\documentclass{aastex}
\usepackage{emulateapj5}

\shorttitle{Infrared Counterpart of GRS~1758$-$258}
\shortauthors{Rothstein et al.}

\begin{document}

\submitted{Accepted for publication in ApJ Letters}

\title{The Infrared Counterpart of the Microquasar GRS~1758$-$258}

\author{D. M. Rothstein,\altaffilmark{1}
S. S. Eikenberry,\altaffilmark{1} S. Chatterjee,\altaffilmark{1}
E. Egami,\altaffilmark{2} S. G. Djorgovski,\altaffilmark{3} and
W.A. Heindl\altaffilmark{4}}

\altaffiltext{1}{Astronomy Department, Cornell University, Ithaca, NY
14853, droth@astro.cornell.edu} \altaffiltext{2}{Steward Observatory,
University of Arizona, 933 North Cherry Avenue, Tucson, AZ 85721}
\altaffiltext{3}{Palomar Observatory, California Institute of
Technology, MS 105-24, Pasadena, CA 91125} \altaffiltext{4}{Center for
Astrophysics and Space Sciences, Code 0424, 9500 Gilman Drive,
University of California, San Diego, La Jolla, CA 92093}

\begin{abstract}

	We present revised infrared ($2.2 \micron$) astrometry of the
field containing the Galactic microquasar GRS~1758$-$258, using
observations at the Keck I 10-m telescope.  We find three candidates
for the microquasar within a $3\sigma$ error circle, but none within
$2\sigma$.  We show that if the 18.4 day X-ray period of
GRS~1758$-$258 is due to a binary orbit, then only one of the three
candidates, an early K-type giant, is large enough to power the
microquasar via Roche lobe overflow.  We therefore identify this star
as the infrared counterpart of GRS~1758$-$258, which we classify as a
low mass X-ray binary.  Long term infrared monitoring of this source
should provide further information about the microquasar system,
including a confirmation of the X-ray period and an estimate of the
compact object's mass.

\end{abstract}


\section{Introduction}

	GRS~1758$-$258 was one of the first objects in the Galaxy to
be identified as a ``microquasar'' $-$ an X-ray binary with jets whose
behavior mimics quasars on a smaller and closer scale.  Because
timescales in microquasars are a factor of $\sim 10^8$ shorter than
quasars, microquasars are excellent laboratories for investigating the
jet formation process in compact objects.

	The status of GRS~1758$-$258 as a microquasar was cemented by
the discovery of radio lobes indicating the presence of collimated jet
outflows \citep{Luis}.  It is one of the two brightest hard X-ray
sources near the Galactic center at energies greater than 50 keV
\citep{Sunyaev}, and its hard X-ray spectra and variability are
similar to that of Cyg X-1 \citep{Kuz}, a black hole candidate which
accretes via the stellar wind from its O-supergiant companion.

	However, the behavior of GRS~1758$-$258 during the transition
between hard and soft spectral states is markedly different from that
of Cyg X-1: there is an observed time delay of $\sim 1$ month between
changes in its luminosity and spectral hardness.  This suggests that
GRS~1758$-$258 has two separate accretion flows, a thin disk and a
halo, and that the observed time delay is equal to the viscous
timescale of the thin disk.  In this case, the calculated disk radius
($3\times10^{10}$ cm) is too large for a wind-fed system and instead
indicates that GRS~1758$-$258 has a low mass companion and is powered
by Roche lobe overflow \citep{Main,SmithA,SmithB}.

	Because of the high optical obscuration ($A_V \sim 10$) along
the line of sight to GRS~1758$-$258, searches for companions have been
mostly restricted to the infrared.  Early attempts did not find any
candidates, but massive supergiants were quickly ruled out
\citep{chen94,Mereghetti}.  \citet{Marti98} found three obscured
optical stars in or near the radio error circle, but due to the
crowded nature of the field, none of the candidates could be
definitively identified as the microquasar counterpart.

	We \citep{Eiken01} obtained deep $K_s$ ($2.15 \micron$) images
of the field and found a $\sim 1.5 \arcsec$ offset between our derived
position for GRS~1758$-$258 and the one obtained by \citet{Marti98}.
Here, we report an error in our original astrometry and obtain a
revised infrared position of GRS~1758$-$258 consistent with that of
\citet{Marti98}.  The apparent $18.45 \pm 0.10$ day X-ray orbital
period of the microquasar \citep{SmithC}, combined with the
requirement that the system undergo Roche lobe overflow, imposes a
constraint on the size of the companion which allows us to identify
one of the stars first seen by \citet{Marti98} $-$ an early K-type
giant referred to as ``Star A'' in \citet{Eiken01} $-$ as the infrared
counterpart of GRS~1758$-$258.

\section{Observations and Data Reduction}

	We present a brief description of the observations, which are
discussed in more detail by \citet{Eiken01}.

	We observed the field containing GRS~1758$-$258 for a total of
94 minutes on June 1, 1998 UTC and 99 minutes on June 2 with the Keck
I telescope, using a $K_s$ filter (centered at $2.15 \micron$) on the
Keck Observatory's Near-Infrared Camera \citep[NIRC;][]{NIRC}.  NIRC
has a 256x256-pixel InSb array with a 0.15 $\arcsec/{\rm pixel}$ plate
scale, giving a $38 \arcsec$ field of view.  The seeing conditions
were exceptionally good, ranging from $\sim 0.35 \arcsec$ to $\sim
0.55 \arcsec$, with extended periods of seeing $< 0.5 \arcsec$.  All
images obtained were shifted and combined to form a master image of
the GRS~1758$-$258 field.

	In order to accurately calibrate the astrometric reference
frame, we obtained wide-field ($8 \arcmin \times 8 \arcmin$) CCD
images of this field in the I-band on June 29, 1998 using the Palomar
1.5-m (60 inch) telescope.  The total exposure time was 300 seconds,
and the typical seeing was $\sim 1.2 \arcsec$.

\begin{figure*}
\epsscale{1.05}
\plotone{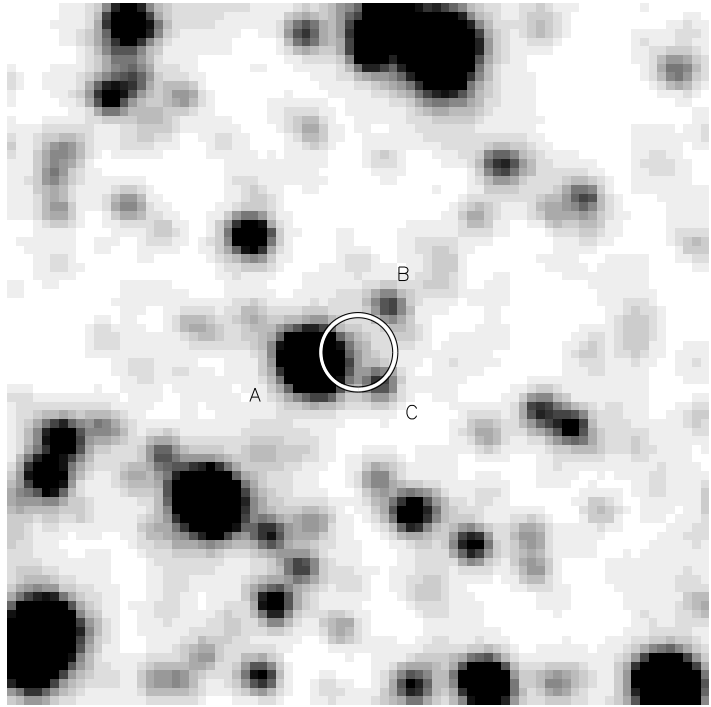}
\caption{\it Close-up ($12 \arcsec \times 12 \arcsec$) Keck $K_s$-band
image of the field of GRS~1758$-$258.  North is up, and east is to the
left.  Our $2\sigma$ error circle ($1.3 \arcsec$ diameter) for the
microquasar position is shown; it is virtually identical to the error
circle obtained in Figure 2 of \citet{Marti98}.  The stars labeled A,
B and C are discussed further in the text.}
\end{figure*}

	We performed astrometry on these images using 9 stars listed
in the USNO-A2.0 astrometric catalog \citep{USNO} that were visible in
the I-band CCD image within $\sim 1.5 \arcmin$ of the radio position
of GRS~1758$-$258 and that did not appear to be double stars, extended
sources or in extremely crowded regions.  There were no suitable USNO
stars visible in the $K_s$-band NIRC image itself, so we obtained a
best fit astrometric solution for the CCD image using the above 9 USNO
stars, determined the coordinates of 5 stars near the position of GRS
1758$-$258 which were visible in both the CCD and NIRC images and
thereby derived a secondary astrometric solution for the NIRC image.

	Figure 1 shows a portion of the NIRC image with the resulting
position of GRS~1758$-$258.  We used the VLA position $\alpha(\rm
J2000) = 18^h \ 01^m \ 12^s.395$ and $\delta(\rm J2000) = -25^o \ 44
\arcmin \ 35 \arcsec .90$ \citep{Felix93}, which we have confirmed
through a reanalysis of the VLA data.  This is also consistent with
the VLA position determined by \citet{Marti2002} using more recent
observations and with the X-ray position determined by {\it Chandra}
\citep{Heindl}.

	Our $2\sigma$ error circle of radius $0.64 \arcsec$ was
determined by combining the $0.1 \arcsec$ uncertainty in the radio
position \citep{Felix93}, the $< 0.1 \arcsec$ uncertainty in the
radio-optical frame tie \citep{daSilva} and our calculated RMS
residuals for the CCD astrometric solution ($0.28 \arcsec$) and the
CCD to infrared solution ($0.05 \arcsec$).  We note that the accuracy
of the radio-optical frame tie in this region of the sky may be worse
than the nominal value (because of the paucity of quasars in the
Galactic plane), so our $2\sigma$ error circle may be an
underestimate.  The GRS~1758$-$258 position obtained here is
consistent with that of \citet{Marti98} $-$ the discrepancy reported
in \citet{Eiken01} was due to a transcription error in shifting CCD
images while performing the astrometry for that paper.

	We find no sources within the $2\sigma$ error circle, but
three sources appear within $3\sigma$ $-$ those labeled A, B and C in
Figure 1.  Star A is at a distance of $0.8 \arcsec$ from the GRS
1758$-$258 position, Star B is at $0.9 \arcsec$ and Star C is at $0.7
\arcsec$.  At the center of the GRS~1758$-$258 error circle, the
pixel-to-pixel variation is heavily dominated by background gradients
from the nearby stars, but we estimate an upper limit of $m_{K_{s}}
\lesssim 18$, at the 95\% confidence level, on any star not detected.

	\citet{Eiken01} performed photometry on Stars A, B and C in
the $K_s$ band and estimated their absolute magnitudes assuming an
extinction of $A_{K_{s}} = 0.9$ mag \citep[derived from the neutral
hydrogen column density along the line of sight to GRS
1758$-$258;][]{Mereghetti} and a distance of 8.5 kpc.  They found that
Star A is consistent with an early K-type giant $-$ as proposed by
\citet{Marti98} based on multi-band photometry and near infrared
spectroscopy $-$ and Stars B and C are consistent with early A-type
main sequence stars.

\section{Discussion}

	In the following discussion we assume that the X-ray emission
from GRS~1758$-$258 is powered by accretion via Roche lobe overflow
\citep{Main,SmithA,SmithB} and that its $18.45 \pm 0.10$ day
periodicity \citep{SmithC} is due to a binary orbit.

\begin{figure*}
\epsscale{1.5}
\plotone{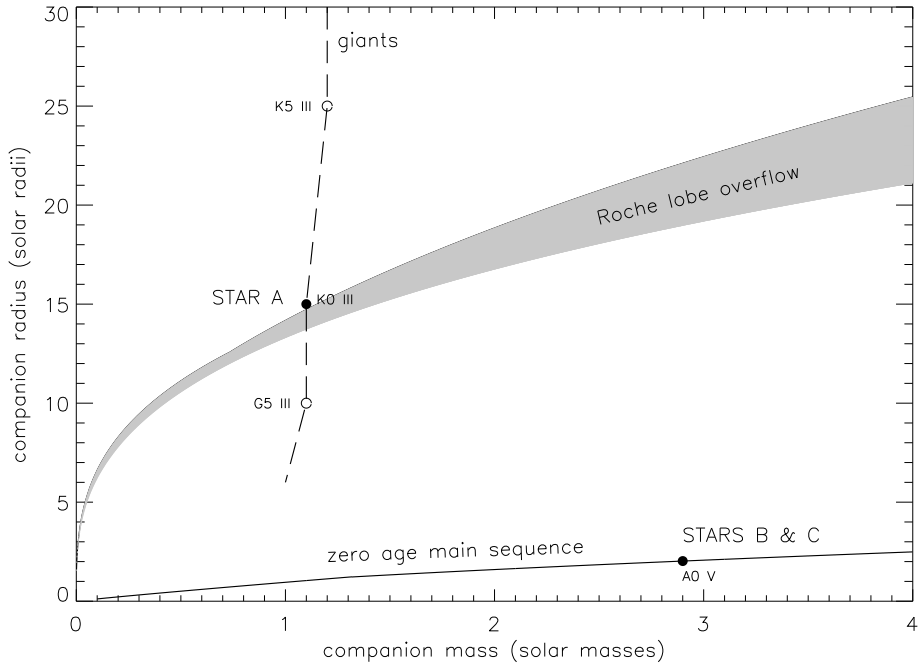}
\caption{\it Possible values of the companion radius for the GRS
1758$-$258 system, as a function of companion mass.  The shaded region
represents the minimum radius required to undergo Roche lobe overflow,
for compact object masses between $1-30 M_{\sun}$ and orbital periods
within $18.45 \pm 0.30$ days.  The solid line shows the mass-radius
relation for zero age main sequence stars, and the dashed line shows
the mass-radius relation for giants, with the approximate locations of
Stars A, B and C overplotted (see text and Figure 1).}
\end{figure*}

	For Roche lobe overflow to occur, the radius of the companion
star must satisfy
\begin{displaymath}
R_{s} \geq \frac{0.49q^{2/3}}{0.6q^{2/3} + \ln(1 + q^{1/3})} a,
\end{displaymath}
where $q$ is the mass ratio of the companion to the compact object and
$a$ is the binary separation \citep{egg}.  Assuming a circular orbit
and the period given above, we can eliminate $a$ and find the minimum
value of $R_{s}$ in terms of the companion and compact object masses.

The shaded region of Figure 2 shows the minimum values of $R_{s}$
required for Roche lobe overflow, as a function of companion mass, for
compact objects between $1-30 M_{\sun}$ and periods within $3\sigma$
of 18.45 days.  Also shown is the mass-radius relation for the zero
age main sequence and for giants \citep{Allen}, with the approximate
locations of our Stars A, B and C marked for clarity.

It is clear that Star A, the early K-type giant, is the only one
consistent with Roche lobe overflow.  Even in the case of an
elliptical orbit, an eccentricity $\ga 0.9$ would be required for
Stars B or C to undergo brief periods of Roche lobe overflow at their
minimum orbital separation, and there is no evidence in the X-ray
light curve of GRS~1758$-$258 for such intermittent behavior.  Any
other star large enough to undergo Roche lobe overflow in this system
would have been detected within our astrometric error circle if it
were located within the Galaxy $-$ for example, a K0 giant would need
to be further than 60 kpc to fall below our detection limit of
$m_{K_{s}} = 18$ (assuming $A_{K_{s}} = 0.9$).  We therefore identify
Star A as the infrared counterpart of GRS~1758$-$258.

\section{Conclusions}

	We have presented revised infrared ($2.2 \micron$) astrometry
of the field containing the Galactic microquasar GRS~1758$-$258.  We
summarize our results as follows:

\begin{itemize}

\item We find three candidates within a $3\sigma$ error circle of the
microquasar position, which is consistent with the results obtained by
\citet{Marti98}.

\item Assuming an $18.45 \pm 0.10$ orbital period of the
GRS~1758$-$258 system \citep{SmithC}, we calculate the radius of the
companion star required to undergo Roche lobe overflow for a range of
companion and compact object masses.

\item We find that only one of our candidate stars (an early K-type
giant labeled Star A in Figure 1) is consistent with Roche lobe
overflow, and we therefore identify this star as the infrared
counterpart of GRS~1758$-$258.

\end{itemize}

	Finally, we note that there are strong $^{12}{\rm C}^{16}{\rm
O}$ absorption bands (equivalent widths $\sim$ 10~\AA) in the infrared
spectrum of the candidate \citep{Marti98}.  Long term monitoring of
these lines should provide a determination of the GRS~1758$-$258 mass
function.  This would allow us to constrain the mass of the compact
object and determine whether it is a neutron star or, as suspected
from its similarity to Cyg X-1, a black hole.

\acknowledgments

	This paper is based in part on observations obtained at the
W.M. Keck Observatory, which is operated by the California Association
for Research in Astronomy, a scientific partnership among the
California Institute of Technology, the University of California and
the National Aeronautics and Space Administration.  The VLA is a
facility of the National Radio Astronomy Observatory, a facility of
the National Science Foundation operated under cooperative agreement
by Associated Universities, Inc.  The authors thank L. Armus and
R. Gal for help in obtaining the CCD images at Palomar Observatory.
We also thank the staff at Palomar and Keck Observatories for their
expert help during our observing runs.  DMR is supported in part by a
National Science Foundation Graduate Research Fellowship.  SSE and DMR
are supported in part by an NSF CAREER award (NSF-9983830).

\end{document}